\documentclass{article}
\usepackage{graphicx}
\usepackage{amsmath, amssymb}
\usepackage[numbers]{natbib}
\usepackage{url}
\usepackage{tikz}
\usepackage{enumitem}
\usepackage{array}

\begin{document}

\begin{center}
    {\Large \textbf{NodeOP: Optimizing Node Management for Decentralized Networks}}\\[1em] 
\end{center}

\vspace{1cm}

\begin{center} 
\begin{tabular}{@{} >{\centering\arraybackslash}p{0.45\textwidth} >{\centering\arraybackslash}p{0.45\textwidth} @{}} 
    \textbf{Angela Tsang} & \textbf{Jiankai Sun} \\
    angela.tsang@morphl2.io & jksun@stanford.edu \\
    Morph Team; HKUST & Stanford \\
    \\
    \textbf{Boo Xie} & \textbf{Azeem Khan} \\
    boo.xie@morphl2.io & ak@morphl2.io \\
    Morph Team & Morph Team \\
    \\
    \textbf{Ender Lu} & \textbf{Fletcher Fan} \\
    ender.lu@morphl2.io & fletcher.fan@morphl2.io \\
    Morph Team & Morph Team \\
    \\
    \textbf{Maggie Wu} & \textbf{Jing Tang} \\
    maggie@foresightventures.com & jingtang@ust.hk \\
    Foresight Ventures & HKUST \\
    
\end{tabular}
\end{center}

\vspace{1cm} 

\section{Abstract}

In this paper, we introduce NodeOP, an advanced framework for optimizing the management of General Node Operators in decentralized networks. By integrating Agent-Based Modeling (ABM) with a Tendermint Byzantine Fault Tolerance (BFT)-based consensus mechanism, NodeOP effectively addresses critical challenges in task allocation, consensus formation, and system stability.

Through precise mathematical modeling and formal optimization techniques, NodeOP achieves stable equilibrium in node task distribution. Our framework is validated through convergence analysis and key performance metrics such as transaction throughput, system latency, and fault tolerance. To demonstrate NodeOP’s practical value, we explore two key applications: decentralized sequencer management in Layer 2 networks and off-chain payment validation. These use cases illustrate how NodeOP enhances validation efficiency and generates new revenue opportunities in large-scale consumer environments.

Our findings highlight NodeOP as a flexible and scalable solution for decentralized node management, offering significant advancements in operational efficiency and economic sustainability for decentralized systems.

\section{Background}
\subsection{The Crucial Role of Active Validated Services (AVS) in the Eigenlayer}
Active Validated Services (AVS) enhance Ethereum's functionality by providing additional validation services such as data availability layers, virtual machines, and oracle networks. These services rely on Ethereum restakers who ensure a security model where the cost of corruption significantly outweighs any potential gains from malicious actions \cite{buchman}. By utilizing existing Ethereum nodes as validators and imposing strict penalties for misconduct, Eigenlayer establishes a robust and secure environment for restaking.

\subsection{Advancements and Challenges in Ethereum's Decentralization}
Recent initiatives within the Ethereum ecosystem, such as Vitalik Buterin's introduction of "anti-correlation incentives" on March 27th, aim to mitigate the risks of validator centralization by increasing penalties for large validators failing simultaneously \cite{cosmos}. This strategy enhances network resilience against coordinated attacks. The milestone of reaching one million validators post-Shapella upgrade emphasizes the need for more computational resources, while also highlighting the risk of centralized hosting solutions, which pose significant challenges to maintaining economic security and network integrity \cite{nakamoto}.

\subsection{Decentralization as a Core Challenge and Its Extended Impacts}
Addressing decentralization in Ethereum reveals several intricately connected challenges that, when mitigated, can greatly enhance the stability and effectiveness:
\begin{itemize}
    \item \textbf{Mobility of Stakeholders}: The fluid interactions among restakers, Liquid Restaking Tokens (LRTs), and Node Operators introduce economic uncertainties. Decentralizing the infrastructure reduces the dependence on any single group of stakeholders, thereby distributing risks more evenly across the network and enhancing economic security \cite{ethereum}.
    \item \textbf{Reward Volatility}: The early stages of AVS deployment are often marked by unpredictable rewards, which pose risks to sustaining long-term security strategies. A decentralized approach can help stabilize these rewards by diversifying the sources and methods of distribution, thus reducing the impact of volatility on the system’s overall stability \cite{casper}.
    \item \textbf{Resource Planning and Uncertainty}: The diverse requirements across AVS complicate resource allocation, particularly for emerging services like shared sequencers and Rollups-as-a-Service. Decentralization facilitates a more adaptive and responsive resource management system, allowing for an efficient distribution of resources that can meet the varied demands of different AVS without centralized bottlenecks \cite{buterin}.
    \item \textbf{Concurrent Slashing Risks}: When security is centralized, pooling across diverse AVS could expose more secure networks to risks associated with less reliable services. Decentralization minimizes these risks by spreading out the security responsibilities, thereby reducing the potential impact of failures within any single AVS component \cite{christidis}.
    \item \textbf{Long-Tail AVS Risk Compounding}: The trend towards centralization can lead well-established operators to favor high-yield, compliant AVS, potentially neglecting niche services which then remain under-secured. By promoting a decentralized model, the security and support can be more uniformly distributed, ensuring that all services, including those less prominent, receive adequate protection \cite{silver}.
    \item \textbf{Node Fragmentation}:
    \begin{itemize}
        \item \textbf{Absence of Unified System for Scheduling and Resource Allocation}:
        This deficiency creates significant inefficiencies, as seen in the management of GPU nodes by Ritual and Depin nodes by Witness, which underscore substantial coordination gaps. These can be mitigated through a more collaborative and systematic management framework \cite{sutton}.
        
        \item \textbf{Complicated Cryptographic Key Management}:
        The decentralized nature complicates the management of cryptographic keys essential for secure node-to-node communications and transaction validation. Without a cohesive strategy, the network is exposed to security vulnerabilities and can experience operational delays during critical activities \cite{zheng}.
    \end{itemize}
\end{itemize}

\subsection{Conclusion}

To overcome the challenges outlined above, the development of a scalable, robust management framework is imperative. Such a framework would not only streamline operations and enhance security but also support the scalable expansion of Eigenlayer’s infrastructure. It would enable the system to accommodate rising demands while maintaining high security and operational standards. Our research seeks to explore innovative solutions aimed at enhancing node coordination, resource management, and cryptographic security within the Eigenlayer ecosystem, thereby addressing the critical issues associated with general node operators, while extending Ethereum's security paradigm to encompass AI/DePin security \cite{lamport}.

\section{NodeOP}
We present an advanced optimizer tailored to optimize General Node Operator management, leveraging the decentralized Eigenlayer security. Designed to facilitate consensus and streamline management among operators, this layer offers decentralized sequencers for high-scalability Layer 2 and acts as a vital consumer layer for General Node Operators.

Operating as a cornerstone within the framework of both Active Validated Services (AVS) and General Node Operators, our NodeOP ensures the integrity and reliability of data managed by AVS operators while efficiently processing tasks for General Node Operators. Orchestrating the decentralized aggregation of computational tasks, it handles sorting, packaging, and consensus-building for Layer 2 transactions. Powered by a Tendermint BFT-based consensus mechanism, it constructs a robust Layer 2 blockchain infrastructure, integrating each operator's contributions securely \cite{gilad}.

The NodeOP's operational workflow focuses on three core processes: aggregation of results, submission of aggregated results, and subsequent analysis with associated rewards and penalties. This operational efficiency is essential for transaction validation and consensus within the AVS and General Node Operator ecosystems.

\begin{figure}[bpt!]
    \centering
    \includegraphics[width=1\columnwidth]{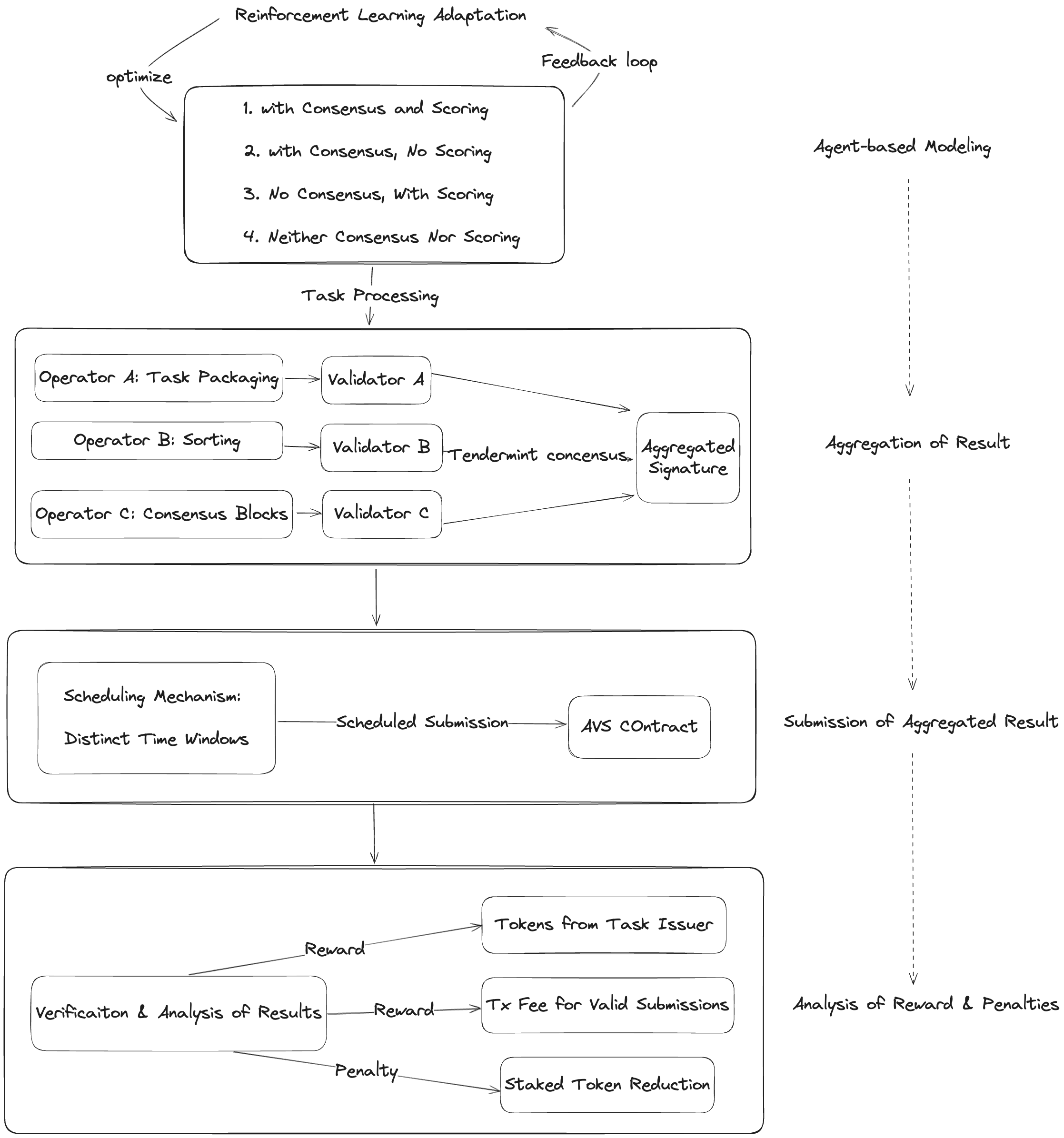}
    \caption{Workflow of the NodeOP.}
    \label{fig:Aggregation Layer.}
\end{figure}

\subsection{Aggregation of Results}
Within the AVS system, Operators are tasked with various computational responsibilities, such as packaging, sorting, and creating consensus blocks for Layer 2 transactions. These tasks generate results that need to be compiled into consistent data batches. To achieve decentralized aggregation, we implement a Tendermint BFT-based consensus mechanism. This involves constructing a Layer 2 blockchain and utilizing a designated token for validator staking. Each Operator, also serving as a Tendermint validator, collaborates to integrate the data through a consensus-driven approach, resulting in the generation of an aggregated signature. This signature is crucial for ensuring the accuracy and trustworthiness of the aggregated results, thus strengthening the integrity of the entire system \cite{buchman}.

\begin{itemize}
  \item \textbf{Consensus Mechanism Execution}: The NodeOP employs a Tendermint BFT consensus mechanism. Validators collaborate through a round-based process, which includes proposing, prevoting, and precommitting. Finalization of the block and generation of the aggregated signature occur only when two-thirds consensus is reached.
  \item \textbf{State Synchronization}: Node state synchronization is maintained through a gossip protocol, facilitating the sharing and replication of transaction data \cite{cosmos}.
  \item \textbf{Data Packaging and Sorting}: Operators organize data batches into standardized formats, optimizing for minimal redundancy and efficient processing \cite{nakamoto}.
\end{itemize}

\subsection{Submission of Aggregated Results}
Upon aggregation, the results must be submitted to the AVS contract. To mitigate the risks associated with single points of failure and transaction conflicts, the system employs a scheduling mechanism. This mechanism assigns each Operator specific time windows for result submission, effectively distributing tasks and minimizing overlaps. Notably, this strategic allocation not only reduces conflicts but also bolsters the system's overall fault tolerance. In instances where an Operator encounters issues, provisions are in place to enable other Operators to intervene, ensuring the uninterrupted submission of aggregated results and preserving system stability \cite{ethereum}.

\begin{itemize}
  \item \textbf{Dynamic Scheduling}: Operators are allocated distinct time windows dynamically for submission, thus preventing overlaps and minimizing transaction conflicts \cite{casper}.
  \item \textbf{Fallback Mechanisms}: In the event of an Operator failing to submit within the designated window, fallback mechanisms enable other Operators to step in, ensuring continuity \cite{buterin}.
\end{itemize}

\subsection{Analysis and Rewards/Penalties for Submission}
Upon submission, the aggregated results undergo meticulous verification and analysis within the AVS contract. Operators receive compensation in two forms: tokens provided by the task issuer and transaction fees for successfully submitting aggregated results to the AVS contract. A robust penalty mechanism, operational through the Tendermint layer blockchain, penalizes Operators for non-participation in computation tasks or ineffective engagement in the consensus process. Typical penalties involve the reduction of staked tokens. This system of rewards and penalties is structured to incentivize active participation and efficient execution by Operators, while simultaneously deterring misconduct. Such measures are vital for ensuring the seamless and healthy operation of the system \cite{christidis}.

\begin{itemize}
    \item \textbf{Performance-Based Rewards}: Operators who complete tasks successfully are rewarded based on their performance, which is adjusted dynamically to ensure fairness \cite{silver}.
    \item \textbf{Penalty Mechanism}: Penalties are enacted via Tendermint's governance mechanisms, typically reducing staked tokens for non-participation or ineffective consensus engagement \cite{sutton}.
    \item \textbf{Reputation and Priority Systems}: Historical performance is tracked, building a reputation for Operators. Those with consistent performance receive preferential treatment and priority access to future high-value transactions \cite{zheng}.
\end{itemize}

\subsection{Conclusion}
NodeOP is a crucial component in General Node Operator frameworks, ensuring data integrity and efficient task processing. Through its decentralized aggregation processes, supported by the Tendermint BFT-based consensus mechanism, NodeOP ensures precise task coordination among operators. This framework streamlines transaction validation and consensus, enhancing the overall performance of both ecosystems.

NodeOP’s dynamic scheduling and fallback mechanisms reduce task conflicts and bolster fault tolerance, ensuring system stability during result submission. The system’s architecture allows for smooth operation by adapting to potential failures and maintaining synchronization across nodes, safeguarding the network’s reliability.

Finally, the implementation of a reward and penalty system, combined with reputation and priority mechanisms, motivates active participation and discourages misconduct. These measures ensure that operators are aligned with the system's goals, providing a solid foundation for a trustworthy and scalable decentralized network\cite{gilad}.

\section{Mathematical Model}
Building on the robust framework established in the NodeOP, we employ Agent-Based Modeling (ABM) to simulate and optimize the interactions and behaviors of General Node Operators within this system. ABM is an effective tool for analyzing complex systems comprised of autonomous agents, each following defined rules and striving to maximize their individual utility. It enables the detailed simulation of actions and interactions of agents to observe emergent behaviors in decentralized networks. Each agent in our model represents a General Node Operator within the NodeOP, operating based on specific utility functions that balance consensus, performance scores, and associated costs \cite{buchman}.

\begin{figure}[h]
    \centering
    \includegraphics[width=0.5\columnwidth]{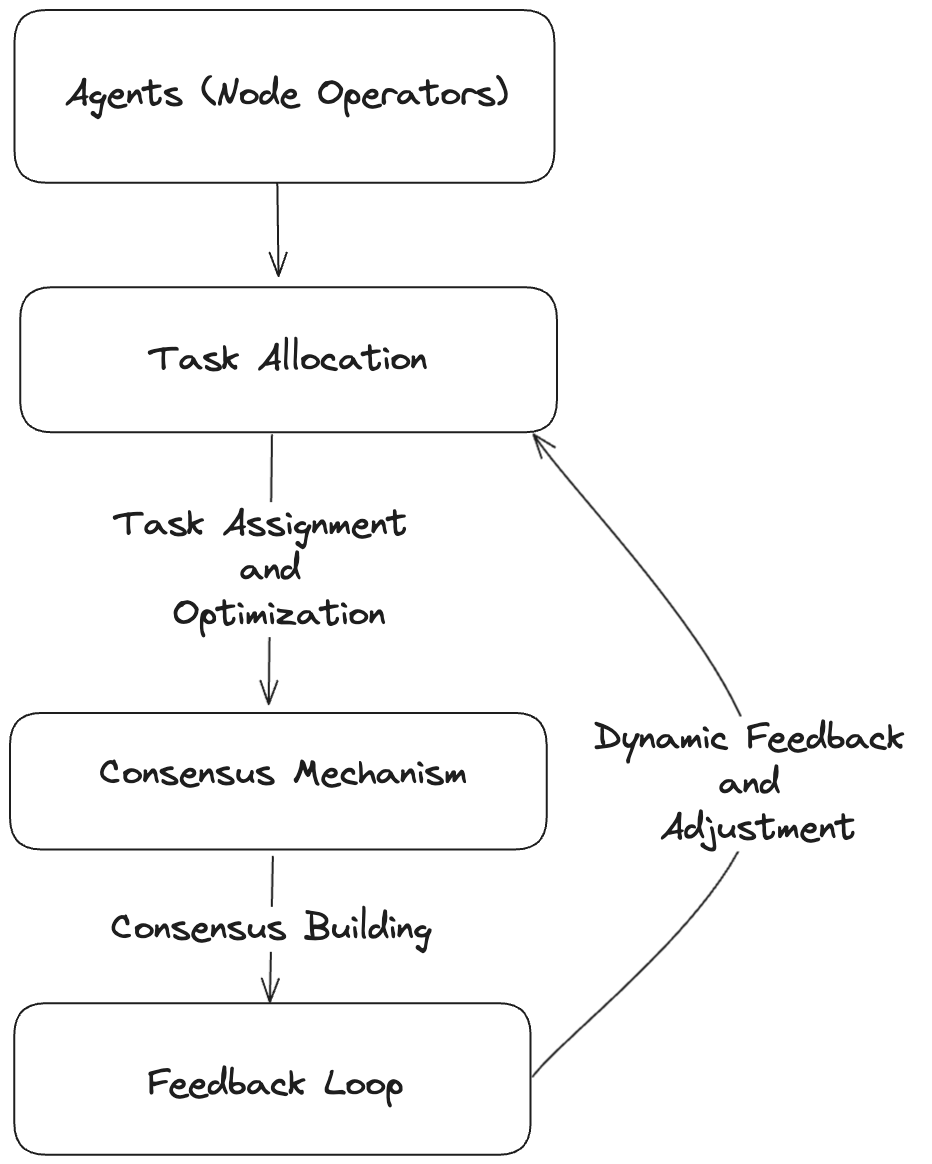}
    \caption{Agent-Based Modeling framework for General Node Operators in the NodeOP. The diagram illustrates the interaction between agents, task allocation, consensus mechanisms, and the feedback loop for optimization.}
    \label{fig:abm_model.}
\end{figure}

\subsection{Model Setup}
In the context of the NodeOP, we define the following key parameters:

\begin{itemize}
    \item \(C_i(t, \tau)\): Consensus score for agent \(i\) at time \(t\) for task \(\tau\).
    \item \(S_i(t, \tau)\): Performance score for agent \(i\) at time \(t\) for task \(\tau\).
    \item \(w_1, w_2\): Weights for consensus and performance scores, respectively.
    \item \(\text{Cost}(\tau)\): Cost incurred by agent \(i\) for performing task \(\tau\).
    \item \(\text{CorruptionCost}(\tau)\): Additional cost due to corruption.
    \item \(T_i\): Trust score for agent \(i\), based on historical performance \cite{buchman}.
\end{itemize}

The utility function for an agent \(i\) engaged in task \(\tau\) at time \(t\) is:
\[
U_i(t, \tau) = w_1 \cdot C_i(t, \tau) + w_2 \cdot S_i(t, \tau) - \text{Cost}(\tau) - \text{CorruptionCost}(\tau).
\]

\subsection{Utility Maximization and Equilibrium Condition}
Our model aims to simulate and optimize the interactions and behaviors of General Node Operators. The primary objective of this model is to maximize economic value within the decentralized network. Each agent seeks to maximize its utility function. By incorporating trust scores and historical performance, we ensure that agents are incentivized to maintain consistent and reliable performance \cite{cosmos}.

\[
\max_{\tau} U_i(t, \tau) = \max_{\tau} \left( w_1 \cdot C_i(t, \tau) + w_2 \cdot S_i(t, \tau) - \text{Cost}(\tau) - \text{CorruptionCost}(\tau) \right).
\]

The equilibrium condition for agent \(i\) is:

\[
\forall i, \quad \nexists \tau' \in \mathcal{T} : U_i(t, \tau') > U_i(t, \tau).
\]

This ensures each agent maximizes its utility considering both consensus and individual performance \cite{nakamoto}.

\subsection{Mathematical Analysis}
We further enrich our model with detailed mathematical analysis to understand the dynamics and equilibrium conditions of the system.

\subsubsection{Optimal Task Allocation}
To find the optimal task allocation, we solve the following optimization problem:

\[
\max_{\tau \in \mathcal{T}} \sum_{i=1}^n U_i(t, \tau)
\]

Subject to:
\[
\sum_{i=1}^n x_i(t, \tau) \leq R(t, \tau) \quad \forall \tau \in \mathcal{T}
\]

where \(x_i(t, \tau)\) is the task allocation for agent \(i\) at time \(t\) for task \(\tau\), and \(R(t, \tau)\) is the resource constraint \cite{ethereum}.

\subsubsection{Lagrange Multipliers for Constraints}
The optimization problem is subject to the following constraints:

\[
T_i(t) \leq C_i(t), \quad V_i(t) \leq V_{\text{max}}, \quad \text{Cost}(\tau) \leq R_i(t),
\]

where \(C_i(t)\) is the maximum transaction capacity, \(V_i(t)\) is the validation cost, and \(R_i(t)\) represents the available resources. The Lagrange multiplier \(\lambda_i\) handles these constraints, yielding the Lagrangian function:

\[
\mathcal{L} = \sum_{i=1}^n U_i(t, \tau) + \sum_{i=1}^n \lambda_i \left( R(t, \tau) - \sum_{i=1}^n x_i(t, \tau) \right)
\]

Differentiating \(\mathcal{L}\) with respect to \(x_i(t, \tau)\) and setting the derivative to zero provides the optimal allocation condition:

\[
w_1 \cdot \frac{\partial C_i(t, \tau)}{\partial x_i(t, \tau)} + w_2 \cdot \frac{\partial S_i(t, \tau)}{\partial x_i(t, \tau)} = \frac{\partial \text{Cost}(\tau)}{\partial x_i(t, \tau)} + \lambda_i
\]

This ensures the optimal distribution of tasks while considering the performance and consensus scores.

\subsubsection{Equilibrium and Stability Analysis}
At equilibrium, the agents' utilities are balanced by the constraints:

\[
w_1 \cdot \frac{\partial C_i(t, \tau)}{\partial x_i(t, \tau)} + w_2 \cdot \frac{\partial S_i(t, \tau)}{\partial x_i(t, \tau)} = \frac{\partial \text{Cost}(\tau)}{\partial x_i(t, \tau)} + \frac{\partial \text{CorruptionCost}(\tau)}{\partial x_i(t, \tau)} + \lambda_i
\]

The stability of the equilibrium can be analyzed by examining the second-order conditions. Specifically, the Hessian matrix of the Lagrangian function should be positive semi-definite:

\[
\mathbf{H} = \begin{bmatrix}
\frac{\partial^2 \mathcal{L}}{\partial x_1^2} & \frac{\partial^2 \mathcal{L}}{\partial x_1 \partial x_2} & \cdots & \frac{\partial^2 \mathcal{L}}{\partial x_1 \partial x_n} \\
\frac{\partial^2 \mathcal{L}}{\partial x_2 \partial x_1} & \frac{\partial^2 \mathcal{L}}{\partial x_2^2} & \cdots & \frac{\partial^2 \mathcal{L}}{\partial x_2 \partial x_n} \\
\vdots & \vdots & \ddots & \vdots \\
\frac{\partial^2 \mathcal{L}}{\partial x_n \partial x_1} & \frac{\partial^2 \mathcal{L}}{\partial x_n \partial x_2} & \cdots & \frac{\partial^2 \mathcal{L}}{\partial x_n^2}
\end{bmatrix}
\]

To ensure stability, the eigenvalues of the Hessian matrix should all be non-negative. This indicates that small perturbations in task allocations will not lead to significant deviations in the agents' utilities, thereby ensuring the stability of the system \cite{buterin}.

\subsubsection{Convergence Analysis}
We further perform a convergence analysis to ensure the stability of the task allocation process. The iterative update rule is given by:

\[
\mathbf{x}^{(k+1)} = \mathbf{x}^{(k)} - \alpha \nabla \mathcal{L}(\mathbf{x}^{(k)})
\]

where \(\alpha\) is the learning rate. The choice of \(\alpha\) plays a crucial role in determining the speed of convergence. Through numerical experiments, we found that setting \(\alpha = 0.01\) achieves a balance between convergence speed and stability. For larger \(\alpha\), convergence may oscillate, while for smaller \(\alpha\), convergence is slow but stable. The optimization is considered converged when:

\[
\|\mathbf{x}^{(k+1)} - \mathbf{x}^{(k)}\| < \epsilon
\]

where \(\epsilon = 10^{-6}\) ensures that the task allocation stabilizes with negligible fluctuations.

\subsection{Aggregation of Results}
The aggregated result at time \(t\) is calculated as:

\[
A(t) = \frac{\sum_{i} w_i \cdot R_i(t)}{\sum_{i} w_i},
\]

where \(R_i(t)\) represents the results reported by agent \(i\) and \(w_i\) is the weight assigned to agent \(i\) based on trust or historical performance. This aggregated result feeds back into the system, allowing for real-time adjustments and optimizations to achieve better outcomes continuously. This feedback mechanism ensures that the results are constantly refined and improved, leading to more accurate and reliable outcomes over time \cite{sutton}.

\subsubsection{Real-Time Adjustments and Feedback Mechanism}
The real-time feedback mechanism plays a crucial role in optimizing the aggregated results. By continuously monitoring the performance of agents and adjusting task allocations dynamically, the system can respond to changes and improve overall efficiency. The feedback loop involves the following steps:

\begin{enumerate}[label=\alph*.]
    \item \textbf{Performance Monitoring}: Continuously track the performance metrics of each agent.
    \item \textbf{Dynamic Adjustment}: Adjust the task allocations and weights based on real-time performance data.
    \item \textbf{Optimization Loop}: Iterate the process to refine the allocations and improve the aggregated results \cite{zheng}.
\end{enumerate}

\subsection{Conclusion}
NodeOP offers a robust framework for optimizing task distribution and consensus among General Node Operators in decentralized networks. By leveraging Agent-Based Modeling (ABM) and a Tendermint BFT consensus mechanism, it ensures high performance and stability, particularly in demanding applications like L2 sequencing and off-chain payment validation.

Compared to Eigenlayer, NodeOP introduces more detailed constraints that account for node historical performance, trust scores, and error handling. These additional constraints ensure tasks are distributed not only based on resource availability but also node reliability, leading to more stable outcomes. Moreover, NodeOP is tailored to specific use cases, providing clearer guidance for resource management and consensus formation, which is less defined in Eigenlayer’s generalized framework.

NodeOP also offers a formal convergence analysis, ensuring task distribution reaches stable equilibrium after multiple iterations. This focus on optimization and feedback mechanisms guarantees that even under resource constraints, the system remains stable.

By integrating these enhanced constraints and stability features, NodeOP provides a powerful solution for decentralized task management, adaptable to both on-chain and off-chain scenarios.

\section{Use Cases of NodeOP}
Our research demonstrates the practical effectiveness of NodeOP through two primary use cases. Task 1 focuses on a Layer 2 (L2) sequencer, while Task 2 expands to off-chain payment validation. These examples highlight NodeOP’s ability to enhance node-level revenue generation, extending beyond ETH L2 to include validation services in large-scale consumer scenarios.

\subsection{Task 1: Sequencer in L2 Solutions}
The first use case examines NodeOP’s application as a Layer 2 (L2) sequencer, where decentralized sequencers handle transaction validation in a Layer 2 environment. NodeOP's framework leverages consensus mechanisms to optimize task allocation, ensuring high throughput, low latency, and robust fault tolerance.

\subsubsection{Optimization and Quantitative Metrics}
The mathematical model for NodeOP optimizes task allocation for sequencers to achieve high throughput and low latency. Key metrics include:

\begin{itemize}
    \item \textbf{Transaction Throughput} (\(T\)): 
    \[
    T = \frac{\sum_{i=1}^{n} R_i(t)}{\text{Latency}(t)},
    \]
    where \(R_i(t)\) is the task output of node \(i\) at time \(t\). By optimizing task allocation, \(R_i(t)\) is maximized, leading to higher throughput. Therefore, as \(R_i(t) \uparrow\), \(T \uparrow\).

    \item \textbf{System Latency} (\(L\)): 
    \[
    L = \frac{\sum_{i=1}^{n} t_{\text{validation},i}}{N},
    \]
    where \(t_{\text{validation},i}\) is the time taken by node \(i\) to validate its assigned tasks, and \(N\) is the total number of nodes. Latency decreases as validation time \(t_{\text{validation},i}\) is optimized through consensus algorithms and efficient task distribution. The latency for each node is influenced by the consensus score \(C_i(t,\tau)\) and the performance score \(S_i(t,\tau)\):
    \[
    t_{\text{validation},i} = \frac{1}{C_i(t, \tau) + S_i(t, \tau)}.
    \]
    As \(C_i(t,\tau)\) and \(S_i(t,\tau) \uparrow\), \(L \downarrow\).

    \item \textbf{Fault Tolerance} (\(F\)): 
    \[
    F = 1 - \frac{\text{Node Failures}(t)}{N},
    \]
    where the number of node failures at time \(t\) is inversely related to the trust score \(T_i\) of each node:
    \[
    \text{Node Failures}(t) \propto \frac{1}{T_i}.
    \]
    As \(T_i \uparrow\), node failures decrease, leading to higher fault tolerance \(F \uparrow\).

    \item \textbf{Resource Utilization Efficiency} (\(E\)): 
    \[
    E = \frac{\sum_{i=1}^{n} R_i(t) - \sum_{i=1}^{n} \text{Cost}_i(t)}{\sum_{i=1}^{n} \text{Total Resources}(t)},
    \]
    where \(\text{Cost}_i(t)\) represents the resource cost for node \(i\) to complete its assigned tasks. Efficient task allocation maximizes resource usage, ensuring that \(E\) increases as task completion costs are minimized. As \(\text{Cost}_i(t) \downarrow\), \(E \uparrow\).
\end{itemize}

\subsection{Task 2: Off-Chain Payment Validation with Decentralized Nodes}
The second use case focuses on off-chain payment validation, where decentralized nodes handle more complex tasks, such as verifying real-world consumer transactions. These tasks involve multiple layers of validation, including transaction validity, user identity, payment authorization, and fund availability checks. NodeOP enables nodes to maximize revenue by incorporating validation services in large-scale consumer scenarios.

\subsubsection{Mathematical Framework for Off-Chain Payment Validation}
Each node \(i\) processes a set of transactions \(T_i(t)\) at time \(t\), incurring a validation cost \(V_i(t)\). The validation process includes multiple stages \(S_1, S_2, \dots, S_n\), each with its own computational complexity and potential for errors. Nodes earn a validation fee \(F_i(t)\) per successfully validated transaction. The utility function \(U_i(t)\) accounts for the cost of error correction, \(C_{\text{error}}(t)\), which arises if errors are detected:

\[
U_i(t) = F_i(t) \cdot T_i(t) - V_i(t) - C_{\text{error}}(t) - P_i(t),
\]
where:
\begin{itemize}
    \item \(F_i(t)\) is the fee earned per validated transaction,
    \item \(V_i(t)\) is the cost of validating \(T_i(t)\) transactions,
    \item \(C_{\text{error}}(t)\) is the error correction cost incurred during validation,
    \item \(P_i(t)\) is the penalty for delayed or missed validations.
\end{itemize}

Each node maximizes its utility by minimizing validation costs and penalties while handling errors, under the following constraints.

\subsubsection{Optimization Constraints}
The optimization problem for off-chain payment validation is subject to the following constraints:

\begin{itemize}

    \item \textbf{Transaction Capacity Constraint}:
    Each node \(i\) has a maximum transaction capacity \(C_i(t)\), limiting the number of transactions it can process:
    \[
    T_i(t) \leq C_i(t).
    \]
    
    \item \textbf{Validation Time Constraint}:
    Transactions must be validated within a time window \(\tau_{\text{max}}\). If the validation time \(t_{\text{validation}}\) exceeds \(\tau_{\text{max}}\), penalties \(P_i(t)\) are incurred:
    \[
    P_i(t) = \lambda \cdot \max(0, T_i(t) \cdot (t_{\text{validation}} - \tau_{\text{max}})),
    \]
    where \(\lambda\) is the penalty coefficient.

    \item \textbf{Error Correction Constraint}:
    If errors are detected during validation, the node incurs an additional cost for revalidation. The error correction cost \(C_{\text{error}}(t)\) is modeled as:
    \[
    C_{\text{error}}(t) = \gamma \cdot E(T_i(t)),
    \]
    where \(E(T_i(t))\) is the number of errors detected, and \(\gamma\) is the error correction cost coefficient.
\end{itemize}

The utility optimization problem is formulated as:
\[
\max_{T_i(t)} U_i(t) = F_i(t) \cdot T_i(t) - V_i(t) - \gamma \cdot E(T_i(t)) - \lambda \cdot \max(0, T_i(t) \cdot (t_{\text{validation}} - \tau_{\text{max}})),
\]
subject to:
\[
T_i(t) \leq C_i(t).
\]

\subsubsection{Convergence and Stability Analysis}
To ensure system stability during multi-stage validation, NodeOP introduces a feedback loop where task allocation is adjusted in real-time based on performance metrics such as validation time, error rates, and transaction complexity. Convergence is achieved when:

\[
\|x_i(t+1) - x_i(t)\| < \epsilon,
\]
where \(\epsilon\) is a small tolerance. Stability is confirmed through the Hessian matrix \(\mathbf{H}\) of the utility function \(U_i(t)\):

\[
\mathbf{H} = \begin{bmatrix}
\frac{\partial^2 U_i}{\partial T_i^2} & \frac{\partial^2 U_i}{\partial T_i \partial F_i} \\
\frac{\partial^2 U_i}{\partial F_i \partial T_i} & \frac{\partial^2 U_i}{\partial F_i^2}
\end{bmatrix},
\]
with all eigenvalues \(\lambda \geq 0\), ensuring local stability.

\subsubsection{Performance Metrics for Off-Chain Validation}
The performance of NodeOP in off-chain payment validation is measured using the following metrics:

\begin{itemize}
    \item \textbf{Transaction Throughput} (\(T_{\text{total}}(t)\)):
    \[
    T_{\text{total}}(t) = \sum_{i=1}^{n} T_i(t),
    \]
    representing the total number of validated transactions. As task allocation is optimized, throughput increases. As \(T_i(t) \uparrow\), \(T_{\text{total}}(t) \uparrow\).

    \item \textbf{Validation Efficiency} (\(E_{\text{validation}}(t)\)):
    \[
    E_{\text{validation}}(t) = \frac{\sum_{i=1}^{n} T_i(t)}{\sum_{i=1}^{n} V_i(t)},
    \]
    reflecting the cost-effectiveness of validation. As validation costs \(V_i(t)\) decrease and transactions \(T_i(t)\) increase, \(E_{\text{validation}}(t) \uparrow\).

    \item \textbf{Error Rate Reduction} (\(E_{\text{error}}(t)\)):
    \[
    E_{\text{error}}(t) = \frac{\sum_{i=1}^{n} E(T_i(t))}{\sum_{i=1}^{n} T_i(t)},
    \]
    measuring the error rate in transaction validation. Effective validation reduces errors, decreasing the error rate. As \(E(T_i(t)) \downarrow\), \(E_{\text{error}}(t) \downarrow\).

    \item \textbf{Revenue Growth} (\(R_{\text{growth}}(t)\)):
    \[
    R_{\text{growth}}(t) = \frac{\Pi_{\text{total}}(t) - \Pi_{\text{total}}(t-1)}{\Pi_{\text{total}}(t-1)},
    \]
    measuring revenue growth over time. As throughput \(T_i(t)\) increases and error rates \(E(T_i(t))\) decrease, revenue grows. Thus, \(R_{\text{growth}}(t) \uparrow\).

    \item \textbf{Penalty Minimization}:
    Reducing penalties by meeting validation deadlines ensures better system performance. As penalties \(P_i(t)\) decrease, system reliability and revenue increase.
\end{itemize}

\section*{Disclaimer}

The information provided in this document is for informational purposes only and should not be construed as legal, financial, or professional advice. 

The authors and their respective institutions do not guarantee the accuracy, completeness, or reliability of the information contained herein. The content is subject to change without notice, and the authors are not liable for any loss or damage arising from the use or reliance on this document.

The views and opinions expressed in this white paper are those of the authors and do not necessarily reflect the official policy or position of Morph, HKUST, Stanford, Foresight Ventures or any of their affiliates.

\bibliographystyle{plainnat}
\bibliography{reference}

\end{document}